\documentclass[aps,prl,twocolumn,groupedaddress,showpacs]{revtex4}
\usepackage{graphicx}
\usepackage{dcolumn}
\usepackage{bm}
\usepackage{amsmath,amsfonts,amssymb,amsthm,verbatim}
\usepackage{floatflt}
\usepackage[ansinew]{inputenc}
\usepackage[usenames]{pstcol}

\usepackage{epstopdf}
\usepackage{color}
\usepackage{epsfig}

\newcommand{\llangle}{\left\langle}
\newcommand{\rrangle}{\right\rangle}

\begin{document}

\title{Tunneling between edge states in a quantum spin Hall system}
\author{Anders Str\"om and Henrik Johannesson}
\affiliation{Department of Physics, University of Gothenburg, SE 412 96 Gothenburg,
Sweden}

\begin{abstract}
We analyze a quantum spin Hall (QSH) device with a point contact connecting two of its edges. The contact supports a net spin tunneling current that can be probed experimentally via a two-terminal resistance measurement. We find that the low-bias tunneling current and the differential conductance exhibit scaling with voltage and temperature that depend nonlinearly on the strength of the electron-electron interaction.

\end{abstract}
\maketitle

A rapidly growing branch of condensed matter physics draws on the exploration of
topologically nontrivial quantum states. Experimentally realized examples, which are
by now well-understood,
are given by the integer \cite{Thouless} and fractional \cite{WenNiu} quantum Hall states. 
These states defy a classification in terms of the standard 
Ginzburg-Landau theory of symmetry breaking and a local order parameter, but can 
instead be characterized by a topological quantity \cite{Thouless,WenNiu}. 
The importance of being able to identify a phase of quantum matter 
that does not fall under the Ginzburg-Landau paradigm has set off a search for other 
topologically nontrivial
states, analogous to, but distinct from those connected to the quantum Hall effects. 

Some time ago, Kane and Mele $-$ building on work by Haldane \cite{Haldane}
$-$ discussed the possibility of a new type of "topologically ordered" state of
electrons in two dimensions: a {\em quantum spin Hall (QSH) insulator}, proposed to
be realized at low energies in a plane of  graphene due to spin-orbit interactions
\cite{KaneMele1}.  Being a band insulator, a QSH insulator has a charge excitation
gap in the bulk, but at its boundary there are gapless edge states with energies
inside the bulk gap. These states, which come in an odd number of Kramers' doublets,
 are ''helical'' (with clockwise/counterclockwise circling states carrying spin
up/down, or vice versa, depending on the orientation of the effective electric field
that enters the spin-orbit interaction) and are responsible for the intrinsic spin
Hall effect that Murakami {\em et al.} had earlier predicted may occur in bulk
insulators \cite{Murakami}.  Time-reversal invariance implies that the energy
levels of the counter-propagating edge states cross at particular points in the
Brillouin zone. It follows that the spectrum of a QSH insulator cannot be
continuously deformed into that of an ordinary band insulator, which has zero (or
equivalently, an even number of) Kramers' doublets. In this exact sense, a QSH insulator
realizes a topologically nontrivial state of matter  \cite{KaneMele2}.  In
subsequent and independent work, the QSH insulator state was proposed to occur also
in strained semiconductors \cite{BernevigZhang} and in HgTe quantum wells with an
"inverted" electronic gap \cite{BHZ}. An experiment carried out on
quasi-two-dimensional HgTe quantum wells grown by molecular beam epitaxy and
sandwiched between (Hg,Cd)Te barriers has revealed data consistent with helical edge
state transport, suggesting the first observation of the QSH effect \cite{Konig}.
The possibility of dissipationless transport of spin currents along the edges of a
QSH insulator is a tantalizing prospect for future spintronics applications
\cite{IBM}. To make progress, however, a more complete picture of the physics is
required. 

An important issue is to understand the behavior of edge currents in the presence of
a tunneling junction connecting two opposite edges of a QSH bar (FIG. 1). When the
bar is connected to a battery, 
a net spin current can tunnel through the junction, and one would like to know how 
the electron-electron interaction influences its conductance. This is the problem we shall address here.

\normalsize
\begin{figure}[ht]
\centering
\includegraphics[width=6.5cm]{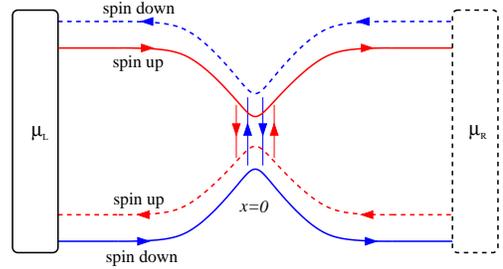}
\centering
\caption{(Color online) Geometry of the QSH point contact device studied in this
paper. The full (dotted) lines represent helical edge states in equilibrium with the
left (right) contact.
}\label{fig:device}
\end{figure}
\normalsize

We consider the simplest situation with a single Kramers' doublet of helical edge
states, applicable to a tunneling experiment on a HgTe quantum well
\cite{Konig}. In the absence of electron interactions this case can formally be
thought of as resulting from a superposition of two integer quantum Hall systems
with the up- and down-spins of the electrons being subject to opposite effective
magnetic fields. This emulates the spin-orbit interaction that is built-in in the
$k$-$p$ Hamiltonian that defines the electron dynamics close to the Fermi level of the
quantum well \cite{Konig}. 
The bar is
connected to a battery with left $(L)$ and right $(R)$ contacts as in FIG. 1.
Applying a gate voltage $V_g$ perpendicular to the upper and lower 
edge of the bar at $x\!=\!0$ will bring the edges close to each other, forming a point 
contact at which electrons may tunnel from one edge to the other.
With the gate voltage turned off there is no tunneling present, assuming the edges
to be well separated.  For the case illustrated in FIG. 1, electrons originating in the $L\  [R]$ contact
of the battery carry current to the right [left], with spin-up [spin-down] on
the upper edge and spin-down [spin-up] on the lower edge. The right- [left-] moving
electrons are in equilibrium with the left [right] contact and have a Fermi
energy equal to the electrochemical potential $\mu_L$ [$\mu_R$] of that contact.
Note that counterpropagating electrons do not equilibrate when injected at different
chemical potentials since any scattering off impurities or defects conserves spin,
thus making impossible transfer of electrons from one type of edge state to the
other. If the driving voltage $V\! \equiv\! (\mu_L\! -\! \mu_R)/e >0$, a net charge
current flows from left to right on each edge, accompanied by a spin current
carrying spin-up [spin-down] on the upper [lower] edge. Neglecting electron
interactions, the ratio of the drain-source charge [spin] edge current to the
driving voltage is the Hall [spin Hall] conductance $e^2/h$ [$e/4\pi$] \cite{footnote}.
When the gate voltage $V_g$ is turned on, more of the right-moving electrons 
tunnel through the point contact at a finite driving voltage $V>0$, leading to a
depletion of the source-drain current. While there is no net tunneling of
charge between the edges, the point contact supports a net inter-edge spin tunneling
current (cf. FIG. 1).

The picture becomes more complex when allowing for electron-electron interactions at
the edges. Away from half-filling of the one-dimensional (1D) band of edge states,
time-reversal invariance constrains the possible scattering processes at an edge to
dispersive $(d)$ and forward $(f)$ scattering \cite{Wu,Xu}. In the vicinity of the
Fermi points  the corresponding interactions are given by
\begin{equation} \label{dispersive}
H_d = g_d\int dx \left(\psi^{\dagger}_{R \uparrow} \psi_{R \uparrow}
\psi^{\dagger}_{L \downarrow} \psi_{L \downarrow}
+ \psi^{\dagger}_{L \uparrow} \psi_{L \uparrow} \psi^{\dagger}_{R \downarrow}
\psi_{R \downarrow}\right)
\end{equation}
and
\begin{equation} \label{forward}
H_f = g_f \sum_{\alpha = R,L \atop \sigma = \uparrow, \downarrow} \int dx \,
\psi^{\dagger}_{\alpha \sigma} \psi_{\alpha \sigma} \psi^{\dagger}_{\alpha \sigma}
\psi_{\alpha \sigma}.
\end{equation}
Here
$\psi_{R \uparrow}$ and $\psi_{L \uparrow}$ are 1D fields that annihilate an
electron in a clockwise propagating helical state on the upper and lower edge,
respectively. Similarly, $\psi_{L\downarrow}$ and $\psi_{R \downarrow}$ are fields
that correspond to a counterclockwise propagating state on the upper and lower edge,
respectively.  It is here important to emphasize that the presence of the dispersive scattering channel, controlled by (\ref{dispersive}), is a fundamental difference between the edge physics of a QSH insulator and a system exhibiting the integer quantum Hall effect (IQHE). As a result, the QSH insulator may show interaction effects which are suppressed in the IQHE. Adding a linearized kinetic term
\begin{equation} \label{kinetic}
H_0 = -iv_F \sum_{\sigma = \uparrow, \downarrow} \int dx \left(\psi^{\dagger}_{R
\sigma}\partial_x\psi_{R \sigma} - \psi^{\dagger}_{L \sigma}\partial_x\psi_{L \sigma}\right), 
\end{equation}
we can bosonize $H=H_0+H_d+H_f$, and obtain
\begin{equation} \label{freeboson}
H = \frac{v}{2}\sum_{i=1,2} \int dx \left(\frac{1}{K}(\partial_x\phi_i)^2 +
K(\partial_x \theta_i)^2\right),
\end{equation}
with $K\!=\!\sqrt{\frac{2\pi v_F+g_f-2g_d}{2\pi v_F+g_f+2g_d}}$, and $v\! =\!\sqrt{( {\small
v_F}\!+\!\frac{g_f}{2\pi})^2\!-\!
(\frac{g_d}{\pi})^2}$, $v_F$ being the Fermi velocity.
The indices 1 and 2 label the upper and lower edge, respectively, with $\phi_1 =
\phi_{R \uparrow}+
\phi_{L \downarrow}, \phi_2 = \phi_{L \uparrow}+ \phi_{R \downarrow}, \theta_1 =
\phi_{R \uparrow}- \phi_{L \downarrow}$, and $\theta_2 = \phi_{L \uparrow}- \phi_{R
\downarrow}$, where $\phi_{\alpha \sigma}$ and $\theta_{\alpha \sigma}$ define
chiral boson fields and their duals within the standard bosonization scheme
\cite{Senechal}. Note that in contrast to an ordinary spinful Luttinger liquid which exhibits spin-charge separation, the
boson fields in (\ref{freeboson}) contain both charge and spin. While helicity makes
spin a redundant quantum number on a single edge, it is important to include it
when two edges are connected via a point contact. The tunneling through the contact,
with amplitude $u$, is governed by the operator 
\begin{equation} \label{tunneling}
H_t  = u \left( \psi^{\dagger}_{L \uparrow} \psi_{R \uparrow} + \psi^{\dagger}_{R
\uparrow} \psi_{L \uparrow} + \psi^{\dagger}_{R \downarrow} \psi_{L \downarrow} +
\psi^{\dagger}_{L \downarrow} \psi_{R \downarrow}\right),
\end{equation}
defined at $x=0$. It can similarly be bosonized:
\begin{equation} \label{bosonizedtunnel}
H_t = \frac{2u}{\pi}\sin[\sqrt{\pi}(\phi_1 + \phi_2)]\cos[\sqrt{\pi}(\theta_1 +
\theta_2)].
\end{equation}
Given the bosonized theory, eqs.\ (\ref{freeboson}) and (\ref{bosonizedtunnel}), we may 
now use standard perturbative RG arguments to
uncover the effect of electron interactions on the tunneling.

As a first step, we integrate out the bosonic fields in the partition function of
the system except at $x=0$, thus obtaining a theory defined only at the location of
the point contact \cite{FN}. With $\Lambda$ an energy cutoff, and $\tau = it$
Euclidean time,  this gives 
\begin{equation} \label{partition}
Z \sim \int \prod_{i=1,2}{\cal D}\phi_i {\cal D}\theta_i \exp(-S -S_t),
\end{equation}
where
\begin{equation} \label{freeaction}
S = \sum_{i=1,2}\int_{-\Lambda}^\Lambda \frac{d \omega}{2\pi} |\omega|\left(
\frac{1}{2K}|\phi_i(\omega)|^2+\frac{K}{2}|\theta_i(\omega)|^2 \right)
\end{equation}
and
\begin{multline} \label{St}
S_t = - \frac{2u}{\pi}\int d\tau \sin\left[
\sqrt{\pi}\left(\phi_1(\tau)+\phi_2(\tau)\right)\right]  \\
\times \cos\left[ \sqrt{\pi}\left(\theta_1(\tau)+\theta_2(\tau)\right)\right].
\end{multline}
Next, the localized fields are split into slow $(s)$ and fast $(f)$ modes,
$\phi_{i s}(\tau)\equiv \int_{-\Lambda/b}^{\Lambda/b}\frac{d
\omega}{2\pi}e^{-i\omega\tau}\phi(\omega)$ and
$\phi_{i f}(\tau)\equiv \int_{\Lambda/b<|\omega|<\Lambda}\frac{d
\omega}{2\pi}e^{-i\omega\tau}\phi(\omega)$,
with $b>1$ a scale factor, and with a similar definition of $\theta_{is}$ and
$\theta_{if}  \ (i=1,2)$.
A cumulant expansion in $u$ then gives an expression for the
low-energy effective action, call it $S_{\textit{eff}}$. To ${\cal O}(u^2)$, 
\begin{equation} \label{effectiveaction}
e^{-S_{\textit{eff}}[\phi_s]}=e^{-S_s[\phi_s]}e^{\left<S_t\right>_f-\frac{1}{2}(\left<S_t^2\right>_f-
\left<S_t\right>_f^2)+\hdots}.
\end{equation}
Here $S_s[\phi_s]$ is the slowly fluctuating part of $S$, while
$\left<\hdots\right>_f$ is an average taken over the fast modes. The calculation of
$\left<S_t\right>_f$ and the second-order cumulant $\left<S_t^2\right>_f-
\left<S_t\right>_f^2$ is here somewhat cumbersome, but is facilitated by the
presence of the time-reversal symmetry. 
We find that
\begin{equation} \label{firstterm}
\left<S_t\right>_f = \frac{2u}{\pi}b^{-{\frac{1}{2}(K+1/K)}} S_t[\phi_{1s},
\phi_{2s}, \theta_{1s}, \theta_{2s}],
\end{equation}
with $S_t[\phi_{1s}, \phi_{2s}, \theta_{1s}, \theta_{2s}]$ as in (\ref{St}) but with
the slow fields replacing the original ones. As for the second-order term,
\begin{multline} \label{secondterm}
\left<S_t^2\right>_f-\left<S_t\right>_f^2 = \frac{u}{\pi}^2\int d\tau \,
( V_\theta \cos\left[\sqrt{\pi}\left( 2\theta_{1} +2\theta_{2}\right)\right]  \\ +
V_\phi  \cos\left[\sqrt{\pi}\left( 2\phi_{1}+2\phi_{2}\right)\right] + \hdots),
\end{multline}
where $V_{\theta}=b^{1-2/K}-b^{1-K-1/K},
V_{\phi}=b^{1-K-1/K} - b^{1-2K}$,
and where $\hdots$ indicate higher-order terms that do not influence the
renormalization to this order in $u$.
The first-order RG equation for $u$, 
\begin{equation}  \label{firstorder}
\frac{du}{d\ln b} = u\left(1-\frac{1}{2}(K+\frac{1}{K})\right),
\end{equation}
is obtained from (\ref{effectiveaction}) and (\ref{firstterm})
and reveals that the scaling dimension $\Delta_K$ of the tunneling operator $H_t$ in
(\ref{bosonizedtunnel}) is $\Delta_K = \frac{1}{2}(K+1/K)$. 
As for the second-order equations, these are extracted from
(\ref{effectiveaction}) and (\ref{secondterm}), and read
\begin{eqnarray}
\frac{d V_\theta}{d \ln b\!}&\!\!=\!\!&\!\frac{u^2}{\pi^2}\left(\!(1\!-\!
\frac{2}{K})e^{(1\!-\! 2/K)\ln b}\!-\!(1\!-\! 2\Delta_K) e^{(1-2\Delta_K)\ln
b}\right), \nonumber \\
\frac{d V_\phi}{d \ln b\!}&\!\!=\!\!&\!\frac{u^2}{\pi^2}\left(\!(1\!-\!
2\Delta_K)e^{(1-2\Delta_K)\ln b}\!-\!
(1\!-\!2K)e^{(1\!-\!2K)\ln b}\right). \nonumber
\end{eqnarray}
These equations imply that, to second order in $u$, $H_t$ renormalizes to zero for
all values of $K$ in the interval $1/2 \!< \!K \!< \!2$. This includes the
experimentally relevant regime for a HgTe quantum well: A rough estimate 
of $K$ for this case, based on the approximate relation $K \approx (1+U/(2E_F))^{-1/2}$
\cite{LiSarmaJoynt}, yields that
$0.8< K < 0.9$, using that $E_F \approx \hbar/m^{\ast} r_s^2$ and $U \approx
e^2/\epsilon\, r_s$, where $e$ is the
electron charge, $\epsilon \approx 20 \epsilon_0$ is the dielectric
constant, $m^{\ast} \!\approx \!0.02 m_e$ (with $m_e$ the electron mass) \cite{Goren}, 
and where $r_s$ is the effective
Bohr radius for electron densities $n_e$ in the interval $0.5\times
10^{11}$ cm$^{-2} < n_e < 3.5\times 10^{11}$ cm$^{-2}$ (at which the experiment in
Ref. \onlinecite{Konig} was carried out).  

It is interesting to compare the
''weak-tunneling'' fixed point found here to the situation for the quantum Hall
effects, where for the IQHE  the tunneling between edge states
is marginal, while for the fractional quantum Hall effect the tunneling renormalizes to
large values for all filling fractions \cite{KaneFisher}. In contrast, as seen from the
second order RG equations above, a
"strong-tunneling" QSH fixed point appears only for $K<1/2$ (or for the unphysical 
region $K>2$ with attractive electron interaction). 

Turning to the tunneling current in the presence of a driving voltage $V = (\mu_L - \mu_R)/e$, we shall focus on the case of a
low bias, allowing us to use a linear response formalism
\cite{Mahan}. The current $I_c(t)$ that we shall calculate is the sum of the charge tunneling currents between edge states with the same helicity, related to the total spin tunneling current $I_s(t)$ by $I_c(t) = (2e/\hbar)  I_s(t)$. Since $I_c(t)$ is equal to the depletion of the charged source-to-drain current in the presence of the point contact (cf. FIG. 1), it follows that the spin current $I_s(t)$ can be detected experimentally by a two-terminal resistance measurement.  

With $V>0$, the current $I_c(t)$ can be expressed as the rate of change of the number of electrons
in equilibrium with the left contact of the battery (see FIG.\ 1), $I_c(t)=-e\langle \dot
N_L(t)\rangle$. The number operator
$N_L= a(\psi^\dag_{R\downarrow}\psi_{R\downarrow}+\psi^\dag_{R \uparrow}\psi_{R
\uparrow})$, where $a$ is a lattice constant, has the property that $\dot N_L=i[H\!+\!H_t,N_L]=i[H_t,N_L]$. This implies that 
\begin{multline}\label{it}
I_c(t)=e u^2 \int d t'\Theta(t-t') \Big(e^{ie\int_t^{t'} dt''V(t'')}\left< [A(t),A^\dag(t')]\right>\\
-e^{-ie\int_t^{t'}dt''V(t'')}\left< [A^\dag(t),A(t')]\right>\Big),
\end{multline}
where
$A= a(\psi^\dag_{L\uparrow}\psi_{R\uparrow}+\psi^\dag_{L\downarrow}\psi_{R\downarrow})$.
Introducing the retarded Green's function
$G_{ret}(t)=-i\Theta(t)\left<[A(t),A^\dag(0)]\right>$ and its transform
$G_{ret}(-eV)=\int dt e^{-ieVt}G_{ret}(t)$, it follows that for constant $V$ the integral in eq.\ (\ref{it}) can be written as  $-2\, \mathrm{Im} [G_{ret}(-eV)]$. The correlation functions 
$G_+(t)=\llangle A(t)A^\dag(0) \rrangle$ and
$G_-(t)=\llangle A^\dag(0)A(t)\rrangle$ are easily calculated in the bosonized theory, and 
one finds that
\begin{equation} \label{korrplusminus}
G_{\pm}(t) = \frac{1}{\pi}\left(\frac{a}{-v(t \pm i\delta)}\right)^{2\Delta_K},
\end{equation}
where $\delta$ is a short-time cutoff. Collecting the
results,
\begin{equation} \label{result}
I_c = 2eu^2 \frac{(a/v)^{2\Delta_K}}{\Gamma(2\Delta_K)}(eV)^{2\Delta_K-1},
\end{equation}
which tells us how the dc tunneling current scales with $V$ in the limit $V\rightarrow 0$, and also
how its amplitude depends on the parameter $K$ that encodes the electron interaction. 
To account for the full $K$-dependence in eq.\ (\ref{result}) one uses
the parameterizations of $K$ and $v$ after eq.\ (\ref{freeboson}), with $g_d \approx
4g_f$ \cite{Senechal}. To ${\cal O}(g_f/v_F)$, $v\approx v_F(5+3K)/(3+5K)$.  

In order to extract the finite-temperature tunneling conductance $G$, we
perform a conformal transformation of the correlation functions in (\ref{korrplusminus}), first
going to Euclidean time $\tau = it$, and then
taking $v\tau \rightarrow
(v\beta/2\pi)\arctan(2\pi \tau/\beta v)$, with $\beta = 1/T$. It follows that
\begin{multline}  \label{Euler}
I_c= -2eu^2(a/v)^{2\Delta_K}(2\pi T)^{2\Delta_K-1}\\
\times \mathrm{Im}\bigg[B(\Delta_K + i eV/2\pi T,\Delta_K - i eV/2\pi T)]\\
\times\frac{\sin\left(\pi(\Delta_K-ieV/2\pi T)\right)}{\cos(\pi \Delta_K)}\bigg],
\end{multline} 
where $B$ is the Euler beta function. In FIG. 2, the current is plotted for a few different values of $K$ and $T$. We have here taken $a \approx 1$ nm 
and $v_F \approx 6 \times 10^6$ m/s \cite{Konig, Goren}, and put $u=0.1 v_F/a$.
From (\ref{Euler}) we obtain the scaling of the zero-bias conductance $G$ with
temperature $T$, 
\begin{equation}
G\equiv \left.\frac{dI_c}{dV}\right|_{V=0} \propto T^{2\Delta_K-2}.
\end{equation} 

\begin{figure}[ht]
\centering
\includegraphics[width=\columnwidth]{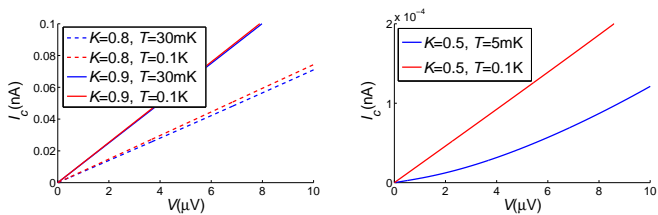}
\centering
\caption{(Color online) The two graphs show the charge tunneling current $I_c$ as  
a function of the applied voltage
$V$\! for different values of $K$ and $T$. 
(The spin tunneling  
current $I_s$ that transfers spin between the edges is given by $I_s  
= (\hbar/2e)I_c$.)
}\label{fig:device}
\end{figure}

It is also interesting to explore the tunneling current for an ac
voltage of the form $V(t)=V_0+V_1\sin(\Omega t)$. Inserting $V(t)$ into eq.\ (\ref{it}) and following Ref. \!\onlinecite{Wen}, we find that the dc component $I_{c,0}$ of the current, defined as the time average of $I_c(t)$,
can be expressed as
$ I_{c,0}=2eu^2 (a/v)^{2\Delta_K} \sum_n
a_n(eV_1/\Omega)(eV_0+n\Omega)^{2\Delta_K-1},$
where 
$a_n(eV_1/\Omega)\!=\!\frac{1}{(2\pi)^2}\int_0^{2\pi}\!\!\int_0^{2\pi}dtdt'e^{in(t'-t)}e^{i\frac{eV_1}{\Omega}(\cos t'-\cos t)}$.
 In FIG.\ 3 we have plotted the dependence of $I_{c,0}$ on $V_0$ for some different values of $K$ and $V_1$. As seen from the figures, $I_{c,0}$ decreases with increasing electron-electron interaction (i.e. with decreasing values of $K$).
 \vspace{26pt}

 \normalsize
\begin{figure}[ht]
\centering
\includegraphics[width=\columnwidth]{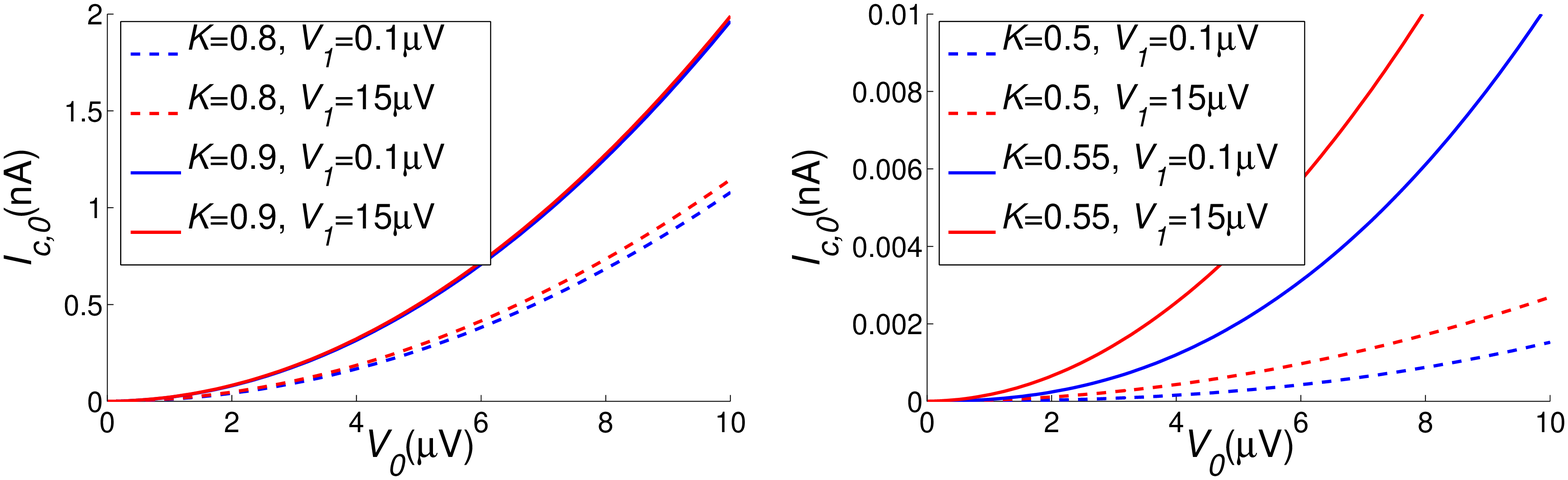}
\centering
\caption{(Color online) The dc component of the charge tunneling current $I_{c, 
0}$ as a function of $V_0$ for different values of $K$ and $V_1$.  
(The dc component of the accompanying spin tunneling current $I_{s,0} 
$ is given by $I_{s,0}= (\hbar/2e)I_{c,0}$.)
}\label{fig:device}
\end{figure}
\normalsize
 
To summarize, we have found that a point contact connecting two edges of a QSH bar supports a spin tunneling current  $I \propto V^{2\Delta_K-1}$ at small voltages $V$, with a zero-bias conductance $G \propto T^{2\Delta_K-2}$ for all values of $\Delta_K = (K+1/K)/2$ with $1/2 < K < 2$, where $K$ encodes the strength of the electron interaction.  This spin current can be probed experimentally via a two-terminal resistance measurement. The interval $1/2 < K < 2$ contains the $K$-values applicable to a HgTe quantum well in the QSH regime \cite{Konig}.  When $K<1/2$, the tunneling amplitude scales to large values, effectively severing the edges, analogous to what happens in a fractional quantum Hall system. Given that a QSH device can be manufactured which allows $K$ to pass through the value of 1/2, this would open for the possibility to experimentally study the transition between strong and weak tunneling in a topologically nontrivial phase of matter. 

We thank A. Brataas, S. Datta and A. Furusaki for valuable correspondence. This work was
supported by the Swedish Research Council under grant 2005-3942. 

{\em Note added:} Upon completion of this work we found a preprint by Hou {\em
et al.}  (arXiv:0808.1723v1, published in Ref.~\onlinecite{Chamon}), on QSH 
edge states in a four-terminal
corner junction geometry. For weak and intermediate electron
interactions these authors find a weak-tunneling fixed point, 
similar to our result for a two-terminal device. 

\end{document}